\documentclass[floats,12pt]{iopart}
\usepackage{iopams}
\usepackage{graphicx}
\usepackage{color}
\newcommand{\bk}{{\bi k}}
\newcommand{\bq}{{\bi q}}
\newcommand{\bQ}{{\bi Q}}
\newcommand{\brr}{{\bi r}}

\begin{document}
\title{Charge Order in the Pseudogap Phase of Cuprate Superconductors} 
\author{W~A~Atkinson,$^1$
  A~P~Kampf,$^2$ and S~Bulut$^{1,2}$} \address{{}$^1$Department of Physics and
  Astronomy, Trent University, Peterborough Ontario, Canada, K9J 7B8} 
  \address{{}$^2$Theoretical Physics III, Center for Electronic Correlations
  and Magnetism, Institute of Physics, University of Augsburg, 86135
  Augsburg, Germany} 
   \eads{\mailto{billatkinson@trentu.ca}, \mailto{Arno.Kampf@physik.uni-augsburg.de}}  
  \date{\today}
\begin{abstract}
Charge ordering instabilities are studied in a multiorbital model of the cuprate
superconductors. A known, key feature of this model is that the large local Coulomb 
interaction in the Cu$d_{x^2-y^2}$ orbitals generates local moments with short range 
antiferromagnetic correlations. The strong simplifying ansatz that these moments are static 
and ordered allows us to explore a regime not generally accessible to weak-coupling approaches.
The antiferromagnetic correlations lead to a pseudogap-like reconstruction of the Fermi 
surface. We find that the leading charge instability within this pseudogap-like state is to a 
phase with a spatially modulated transfer of charge between neighboring oxygen $p_x$ and $p_y$
orbitals accompanied by weak modulations of the charge density on the Cu$d_{x^2-y^2}$ orbitals. 
As a prime result of the antiferromagnetic Fermi-surface reconstruction, the wavevectors of 
the charge modulations are oriented {\em along} the crystalline axes with a periodicity that 
agrees quantitatively with experiments. This suggests a resolution to a discrepancy between 
experiments, which find axial order, and previous theoretical calculations, which find 
modulation wavevectors along the Brillouin zone (BZ) diagonal. The axial order is stabilized 
by hopping processes via the Cu$4s$ orbital, which is commonly not included in model analyses 
of cuprate superconductors. The main implication of our results is that charge order emerges 
{\em from} the pseudogap state, and is not the primary source of the pseudogap.

\end{abstract}
\maketitle


\section{Introduction}
\label{sec:Introduction}
Cuprate high-temperature superconductors are, over a broad range of doping, characterized by
anomalous thermal, transport, and spectral properties \cite{Timusk:1999wp}. These are due to a
``pseudogap'' phase, which has been attributed variously to incoherent fluctuations of 
incipient antiferromagnetic (AF) \cite{Kampf:1990,Chubukov:1997vp,Schmalian:1998hv,Abanov:2003cs,Sedrakyan:2010}, stripe \cite{Kivelson:1998ir}, or superconducting order 
\cite{Randeria1992,Emery:1995dr,Kwon1999,Eckl2002}, or combinations thereof 
\cite{Alvarez2008,Atkinson:2012,Hayward:2014,Meier:2014}; to strong correlation physics 
\cite{Kyung:2006wb} and to exotic microscopic nonsuperconducting phases, including ``loop 
currents" \cite{Varma:1997} and ``$d$-density waves" \cite{Chakravarty:2001}.
Experimentally, the physics underlying the pseudogap has proven difficult to isolate, in part 
because unambiguous signatures for the different pseudogap scenarios are lacking.

Renewed efforts to understand possible connections between the pseudogap, superconductivity, 
and non-superconducting phases have been spurred by evidence for charge order in various 
underdoped cuprates, including YBa$_2$Cu$_3$O$_{6+x}$ (YBCO)
\cite{Wu:2011ke,Ghiringhelli:2012bw,Chang:2012vf}, Bi$_2$Sr$_2$CaCuO$_{8+\delta}$ 
\cite{Hoffman:2002bk,Kohsaka:2007hx}, Bi$_2$Sr$_{2-x}$La$_x$CuO$_{6+\delta}$ (BSCCO)
\cite{CominScience2014}, and HgBa$_2$CuO$_{4+\delta}$ (HBCO) \cite{Doiron:2013,Barisic:2013kz}.  Notably, 
charge order is observed at similar doping levels to where the pseudogap is observed.    
Incommensurate charge modulations oriented along the crystalline axes, with 
wavevectors near $q^*=0.3$ reciprocal lattice units, were seen by resonant x-ray
scattering (RXS) \cite{Ghiringhelli:2012bw,CominScience2014,daSilvaNeto:2014bz,Comin:2014vq},
x-ray diffraction \cite{Chang:2012vf,Blackburn:2013,Achkar:2012ep,Blanco-Canosa:2013},
and scanning tunneling microscopy (STM)
\cite{Kohsaka:2007hx,Wise:2008,Lawler:2010n,CominScience2014,daSilvaNeto:2014bz} in zero 
magnetic field. NMR \cite{Wu:2011ke,Wu:2013} and ultrasound experiments \cite{LeBoeuf:2012up}
found that the charge correlations are long-range only in finite magnetic fields. 
Consistent with the onset of some kind of electronic order, a Fermi surface reconstruction was 
revealed by quantum oscillation experiments \cite{Sebastian:2012wh,Harrison:NJP2014}, and by 
transport measurements of Hall, Seebeck, and Nernst coefficients 
\cite{Chang:2010ic,Chang:2011fw,Doiron:2013}. Ultrasound data suggest that the charge 
modulations form a biaxial ``checkerboard" pattern \cite{LeBoeuf:2012up}, while STM data have 
been interpreted either in terms of checkerboard \cite{Wise:2008} or uniaxial 
\cite{Kohsaka:2007hx,Lawler:2010n,Fujita:2014kg} order. A direct causal connection between charge order and 
the onset of pseudogap features at a temperature $T^\ast$ appears unlikely: first, the charge 
ordering temperature $T_\mathrm{co}$ typically lies below $T^\ast$ 
\cite{Chang:2012vf,ChanarXiv2014}; second, the ordering wavevector $q^\ast$ does not connect 
Fermi surface sections at the BZ boundary from which the pseudogap emerges 
\cite{CominScience2014,daSilvaNeto:2014bz}. Nonetheless, it has been proposed that charge 
ordering fluctuations above $T_\mathrm{co}$ may contribute essentially to the pseudogap 
\cite{Meier:2013,Sachdev:2013}.

An intriguing feature of charge order in YBCO and BSCCO is that there appears to be a strong 
{\em intra}-unit cell transfer of charge between oxygen atoms in each CuO$_2$ plaquette, rather 
than the {\em inter}-unit cell charge transfer normally associated with charge-density waves.  
The most direct evidence for this comes from  STM experiments 
\cite{Kohsaka:2007hx,Lawler:2010n,Fujita:2014kg}, and further support is provided by x-ray scattering
\cite{Comin:2014vq}. Roughly then, the charge ordered phase can be thought of as a finite-$\bq$ 
modulation with a $d_{x^2-y^2}$ form factor describing the intra-unit cell charge transfer,  and with  relatively little transfer of charge between neighbouring unit cells. For 
this reason, the phase is sometimes called a ``$d$CDW''.  Alternatively, because the charge 
order is a generalization of a $\bq= (0,0)$ nematic phase that  breaks rotational but not 
translational symmetry, it has been labelled a ``modulated nematic''. 

It is natural to ask whether the $d$CDW charge order identified in YBCO and BSCCO is related to stripe order (see \cite{Thampy:2013ty,Achkar:2014} for further discussion).  Stripe order is  well established in  La$_{2-x}$Ba$_x$CuO$_4$ and dynamical stripes are inferred in La$_{2-x}$Sr$_x$CuO$_4$\cite{Kivelson:2003}.  Stripes in La$_{2-x}$Ba$_x$CuO$_4$ are characterized by a static or quasistatic spin modulation whose period is double that of a concomitant charge modulation.\cite{Vojta:2009}  The doping dependence of the modulation wavevector is opposite to what one would expect for a Fermi surface instability, and suggests instead a strong coupling picture in which holes and spins segregate into one dimensional stripes\cite{Vojta:2009}.   The $d$CDW described above has some similarities to this stripe order:  both compete with superconductivity, and both have maximal intensity near a hole doping $p=1/8$ in all cuprates for which the doping dependence has been measured.\cite{Huecker:2011,Huecker:2014vc,Blanco-Canosa:2014}  On the other hand, there are also significant differences.  First, the local spins in YBCO are dynamic, rather than (quasi)static. Models of fluctuating stripes have  been proposed to describe this\cite{Kivelson:2003,Vojta:2006,Seibold:2012132}; however, recent NMR experiments clearly show that the charge order is static in  YBCO  up to high temperatures.\cite{Wu:2014vx}  This suggests that the intertwining of charge and spin textures that is key to stripe formation in the La-cuprates is not a factor in YBCO.
 Consistent with this we note that, while
the doping dependences of the spin and charge modulation wavevectors are closely connected in  La$_{2-x}$Ba$_x$CuO$_4$, they appear unconnected in YBCO\cite{Blackburn:2013}.  
 Finally, recent x-ray experiments have shown that the structure factor for charge order in La$_{2-x}$Ba$_x$CuO$_4$ has an extended-$s$ symmetry,\cite{Achkar:2014}  consistent with multiorbital models of a magnetically driven stripe instability,\cite{Lorenzana:2002}  and in contrast to YBCO and BSCCO.
 
  Whether these differences are due to small differences in the band structure that tip the balance towards particular phases, or point to larger differences between the cuprate families is not yet established.  Here, we adopt the point of view that the mechanism driving charge order in YBa$_2$Cu$_3$O$_{6+x}$ is distinct from that in the La-cuprates.
 The majority of previous  theoretical work along these lines is based on one-band effective models of a single 
CuO$_2$ plane; in such models, the analogue of intra-unit cell charge redistribution is bond 
order, namely, an anisotropic renormalization of the electronic effective mass along the $x$ 
and $y$ axes. Several theories have argued that bond order follows from AF exchange
interactions;
a vital role for the charge instabilities is thereby ascribed to ``hot spot'' regions of the 
Fermi surface where scattering from AF spin fluctuations is especially strong \cite{Metlitski:PRB2010,Metlitski:2010vf,Sachdev:2013,Efetov:2013,Meier:2013,Sau:2013vw,Wang:2014wc,Chowdhury:2014}. Alternative 
one-band \cite{Holder:2012ks,Husemann:2012vg,YamasePRB2012} and three-band \cite{Bulut:2013} 
model calculations with generic interactions have found similar charge instabilities. With the 
exception of Ref.~\cite{Wang:2014wc}, which additionally found current-carrying stripes, these 
models universally obtained a charge density with a $d_{x^2-y^2}$ form factor and an ordering 
wavevector $\bq^\ast$ along the BZ diagonal. While the form factor is compatible with 
experiments \cite{Kohsaka:2007hx,Lawler:2010n,Comin:2014vq}, the magnitude of $\bq^\ast$ is 
typically too small by a factor of 2, and the direction of $\bq^\ast$ is rotated by $45^\circ$ 
relative to the experiments. The robustness of these discrepancies suggests that the underlying 
models lack an essential ingredient. 

In this work, we show that the Fermi surface topology affects the emergent charge order in a 
fundamental way, and can explain the discrepancy between the observed and predicted values of 
$\bq^\ast$.  Starting from a simplified model of the  Fermi surface in the pseudogap phase, we 
obtain a charge instability that quantitatively agrees with that found experimentally. The 
implication of this work is that  charge order emerges {\em from} the pseudogap phase, rather 
than contributing to it directly.

Experimentally, the pseudogap is characterized by a partial depletion of the density of states 
around the Fermi level.  Photoemission experiments have revealed that this depletion occurs 
near the BZ boundary at $(\pm \pi,0)$ and $(0,\pm \pi)$, and that spectral weight near these 
points is pushed away from the Fermi energy \cite{Damascelli:2003}. Early proposals ascribed 
this to nearly AF spin fluctuations that partially nest these regions of the Fermi surface and 
shift spectral weight to higher energy \cite{Kampf:1990,Chubukov:1997vp}. This picture remains 
physically appealing because the underdoped regime lies near the AF insulating phase of the  
parent compounds, and it is supported by quantum Monte Carlo (QMC) \cite{Preuss:1997tu} and 
cluster dynamical mean-field theory \cite{Kyung:2006wb} (cDMFT) calculations that draw a link 
between spin fluctuations and the pseudogap.

With this in mind, we adopt a model that we believe contains the essential ingredients to 
understand charge order in the cuprates. The basic element of this model is a two dimensional 
CuO$_2$ plane, and we retain both Cu and O orbitals, as well as the short-range Coulomb 
interactions between them.  While it is likely that similar ordering instabilities to the ones described here may be found in one-band models, by choosing a multiorbital model we are able to obtain details of 
the intra-unit cell charge redistribution, which can be directly probed by STM, NMR, and x-ray 
experiments. We examine specifically the charge instabilities generated by the short-range 
Coulomb forces, which have been shown to be attractive in the relevant charge ordering channel 
\cite{Fischer:2011,Bulut:2013}. We note that spin fluctuations are also attractive in this 
channel, at least in one-band models; these will modify $T_\mathrm{co}$, but should not alter the
relationship between $\bq^\ast$ and the Fermi surface structure. For simplicity, then, we focus 
on the charge fluctuations only and omit spin dynamics.

Pseudogap physics in the three-orbital model derives from the large local Coulomb interaction, 
$U_d \sim 10$ eV, on the Cu sites, which is the source of strong correlation physics in the 
cuprates. Despite the achievements of
state-of-the-art computational methods, there is 
still a paucity of tools available that can capture both the short-range strong correlation 
physics and the long-range physics of incommensurate charge order. Numerical methods like QMC 
and cDMFT, which have proved capable of verifying the pseudogap structures in the density of states, 
suffer from finite-size effects that render the charge instability inaccessible, and cluster 
methods further face the difficulty of treating the nonlocal Coulomb interactions that drive 
the charge order \cite{Aichhorn:2004}. On the other hand, weak-coupling methods, which capture 
long-range physics, find that spin fluctuations introduce only weak pseudogap-like spectral 
features \cite{Chubukov:1997vp,Schmalian:1998hv,Abanov:2003cs,Sedrakyan:2010}.

For these reasons, we follow a partially phenomenological approach. To leading order, the effect
of $U_d$ is to suppress double occupancy of the Cu d$_{x^2-y^2}$-orbital and thereby create local 
moments;  as a consequence, the itinerant electrons reside primarily on the oxygen sites, although the Fermi surface does nevertheless have some Cu character due to the hybridization of Cu and O orbitals.  Above the superconducting transition, the spin spectrum measured by neutron scattering\cite{Bourges:2000fl} is centred at $(\pi,\pi)$, indicating dynamical AF correlations.
Our subsequent diagrammatic analysis is based on two strong simplifying  assumptions: first, the 
moments are assumed quasistatic on electronic timescales, and second, the AF correlation length 
$\xi_\mathrm{AF}$ is larger than the charge order correlation length $\xi_\mathrm{co}$, which is 
estimated from experiments to be $\sim 50$~\AA\cite{Huecker:2014vc}. In essence, this implies
that the local moments can be treated, as if they are ordered antiferromagnetically.   (A similar ansatz was made in \cite{Fischer:2014}, where static spin textures of classical magnetic moments on the copper sites were shown to induce charge order.)  Both of 
these assumptions are not satisfied throughout most of the doping range where charge 
order is observed experimentally ($\xi_\mathrm{AF}\sim 20$ \AA{} in YBa$_2$Cu$_3$O$_{6.5}$\cite{Stock:2004gu}); however, by making these assumptions we are able to explore a different physical regime than previous weak coupling calculations. 
We also note that there is  
evidence in YBa$_2$Cu$_3$O$_{6+x}$ that charge order 
survives inside a static magnetic phase that exists at very low doping.\cite{Huecker:2014vc}  Ultimately, however, one should think of the above approximations as a simple phenomenology for strong correlation physics on the copper sites, which is justified after the fact by the surprising accuracy with which we predict certain properties of the charge ordered phase.

The effect of the ordered local moments is to create a pseudogap-like shift of spectral weight 
away from the Fermi level at $(\pm \pi,0)$ and $(0,\pm \pi)$ and to reconstruct the Fermi 
surface to form hole and electron pockets. In this pseudogapped state we find that residual 
Coulomb interactions between the quasiparticles can drive a $d_{x^2-y^2}$-like charge 
redistribution between O$p_x$ and O$p_y$ orbitals, accompanied by a weaker periodic modulation 
of the Cu charge density. The obtained charge pattern with an ordering wavevector $\bq^\ast$ 
along the BZ axis is indeed consistent with what has been observed experimentally. This charge 
order induces a second Fermi surface reconstruction which generates diamond-shaped electron 
pockets. The existence of such pockets was earlier inferred from quantum-oscillation 
experiments.

We introduce the model in section \ref{sec:calc} and describe briefly the calculations for the 
charge susceptibility, with details left for the appendices. The results of our calculations 
are discussed in section \ref{sec:results}, with an emphasis on comparisons to experiments. The 
main implication is that the proposed model calculation, while still not a complete description 
of the microscopic physics underlying charge order, provides a route to understand the 
experiments, and suggests that charge order and pseudogap features are in fact distinct 
phenomena. A short summary is contained in section \ref{sec:conclusions}.

\section{Model and Calculations}
\label{sec:calc}
The goal is to model charge order in YBa$_2$Cu$_3$O$_{6+x}$, and to this end we employ a 
multiband description of the CuO$_2$ planes due to Andersen {\em et al.} (ALJP) that was derived
specifically for YBa$_2$Cu$_3$O$_{7}$ \cite{Andersen:1995}. In an extension to the Emery model 
\cite{Emery:1987prl}, which is based on the Cu$3d_{x^2-y^2}$ and two $\sigma$-bonded oxygen 
orbitals, O$p_x$ and O$p_y$, ALJP included also the Cu$4s$ orbital. The latter resides well 
above the Fermi energy, approximately 6.5~eV above the Cu$d$ orbital, and has a large overlap 
with the O$p$ orbitals. Downfolding this orbital leads to an effective three-band model 
(see~\ref{sec:A1}), ${\cal H} = \sum_\bk\psi^\dagger_\bk 
{\bf H}(\bk) \psi_\bk$, where
\begin{equation}
{\bf H}(\bk) = \left[ \begin{array}{ccc} 
    \epsilon_{d} & 2t_{pd} s_x & -2t_{pd} s_y \\
    2t_{pd} s_x & \tilde \epsilon_{x}(\bk) &   4\tilde t_{pp} s_x s_y \\
    -2t_{pd} s_y & 4\tilde t_{pp} s_x s_y & \tilde \epsilon_{y}(\bk)
\end{array}
\right ]
\end{equation}
and $\psi^\dagger_\bk = [d^\dagger_\bk, p_{x\bk}^\dagger, p_{y\bk}^\dagger]$ is an array of {\em electron} 
creation operators for the $d$, $p_x$, and $p_y$ orbitals. Parameters $t_{pd}$ and $t_{pp}$ 
denote hopping amplitudes, $s_{x,y}=\sin(k_{x,y}/2)$, $\tilde \epsilon_{x,y}(\bk) = \epsilon_p+
4t^i_{pp}s_{x,y}^2$, and $\epsilon_d$ and $\epsilon_p$ are orbital energies. The tilde denotes 
renormalization by hopping through the Cu$4s$ orbital. In particular, $\tilde t_{pp}=t_{pp}^d + 
t_{pp}^i$ where the superscripts indicate direct ($d$) and indirect ($i$; through the Cu$4s$ 
orbital) hopping between O$p$ orbitals.

\begin{figure}[tb]
\begin{center}
\includegraphics[width=0.5\columnwidth]{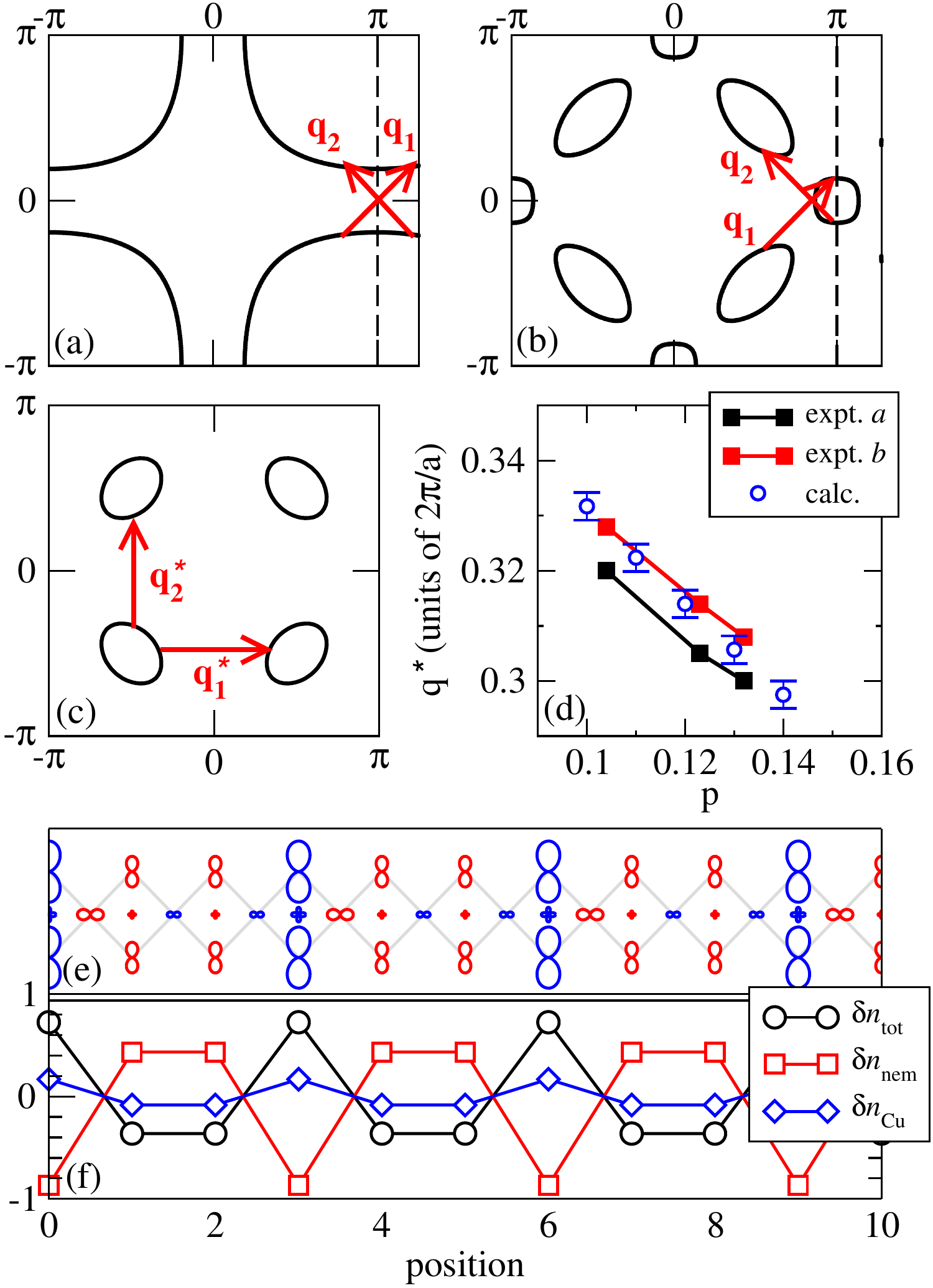}
\end{center}
\caption{Leading charge instabilities. (a) ALJP Fermi surface and the 
  calculated modulation wavevectors ${\bq_1}$ and ${\bq_2}$ at which the charge
  susceptibility first diverges. Fermi surfaces and concomitant charge ordering wavevectors 
  are also shown for (b) $M=0.5$ eV and (c) 
  $M=1.5$ eV. All three 
  figures are at a hole density of $p=0.10$ where $p\equiv 5-n$ and  $n$ is the total electron 
  density. (d) Magnitude of
  the modulation wavevector $|\bq^\ast_1|$ for $M=1.5$ eV as a function of hole density 
  together with experimental results from Ref.~\cite{Blackburn:2013} for YBa$_2$Cu$_3$O$_{6+x}$ 
  along $a$ and $b$ axial directions. Error bars indicate the $q$-resolution of our 
  calculations. The results are for the temperature $T=110$ K (see Fig. 4 for corresponding 
  critical $V_{pp}$ values).
  (e) Orbitally resolved charge modulations for unidirectional charge order and $p=0.10$.   
  The sizes of the Cu$d$, O$p_x$, and O$p_y$ orbitals indicate the relative sizes of the 
  positive (red) and negative (blue) charge modulations on those orbitals. We have taken 
  $q^\ast = 1/3$ for presentation purposes. (f) Modulation of the total charge per unit cell 
  $\delta n_\mathrm{tot}$, nematic modulation (see text) $\delta n_\mathrm{nem}$, and Cu charge 
  density $\delta n_\mathrm{Cu}$. Note that {\em relative} amplitudes are shown. 
  The horizontal axes in (e) and (f) are the same.} 
\label{fig:pdep}
\end{figure}

For the reasons outlined above we introduce AF moments on the Cu$d$ orbitals by adding a 
staggered spin-dependent potential $M(\brr_j)$ to the Hamiltonian and thereby obtain a 
pseudogap-like reconstruction of the Fermi surface. It is natural to think of this potential as
the auxilliary field that appears when the Coulomb interaction $U_d \hat n_{jd\uparrow} 
\hat n_{jd\downarrow}$ on the Cu$d$ orbitals is removed by a Hubbard-Stratonovich transformation.  
In this transformation, the quartic interaction term is replaced by an interaction between the 
electrons and a spin-polarizing time-dependent auxiliary field $M(\brr_j,t)$. As mentioned in 
section~\ref{sec:Introduction}, we make two assumptions in order to isolate the physics of 
interest: first, that the field is static, and second that $M(\brr_j)$ has long range AF order. 
Under these assumptions, an additional term, $-M \sum_j e^{i\bQ\cdot\brr_j} ( \hat n_{j\uparrow} -
\hat n_{j\downarrow} )$ with $\bQ = (\pi,\pi)$, is added to the Hamiltonian. On physical grounds, 
we expect this potential to be quite large: within a saddle-point approximation, $M=U_d m$, 
where $m$ is the static AF moment on the Cu sites. Given that $U_d \sim 10$ eV in the cuprates, 
even a modest value of $m$ leads to $M \sim 1$ eV. The Fermi surface reconstruction generated 
by $M$ is illustrated in Fig.~\ref{fig:pdep}, where the local Cu$d$ moments open a gap along 
Fermi surface segments near the AF hot spots, i.e. those points where the Fermi surface 
intersects the magnetic BZ boundary.

Charge order is driven by interactions between quasiparticles in the reconstructed bands. It 
has been shown that, in one band models at least, the exchange of spin fluctuations may drive a 
charge ordering transition; here, we consider only short range Coulomb interactions. Electrons 
interact at short distances through intra-orbital $U_d$ and $U_p$ and nearest-neighbor $V_{pd}$ 
and $V_{pp}$ Coulomb repulsions.  The corresponding interaction part of the Hamiltonian is
\begin{eqnarray}
\fl \hat V = \sum_j \Big [ U_d \hat n_{jd\uparrow} \hat n_{jd\downarrow}
+ U_p \left ( \hat n_{jx\uparrow} \hat n_{jx\downarrow} + 
\hat n_{jy\uparrow} \hat n_{jy\downarrow} \right ) 
+ V_{pd} \sum_{\delta} \sum_{\alpha=x,y}
\hat n_{jd} \hat n_{j+\delta\, \alpha}  \nonumber \\
+V_{pp} \sum_\delta \hat n_{jx}\hat n_{j+\delta\, y}
\Big ],
\label{eq:coulomb}
\end{eqnarray}
where $\sum_j$ implies summation over unit cells, and $\delta$ is summed over nearest-neighbor 
orbitals of type O$p_{x,y}$ (for $V_{pd}$) or O$p_y$ (for $V_{pp}$).  In our model, the charge instability is driven by $V_{pp}$ .

To study charge ordering tendencies, 
we calculate the charge susceptibility $\chi_{\alpha\beta}(\bq)=-(\partial n_\alpha/\partial 
\epsilon_\beta)(\bq)$, where $n_\alpha$ denotes electron densities and $\alpha$ and $\beta$ 
are orbital labels.  The onset of charge order is signalled by a 
diverging susceptibility at a specific ${\bf q}^*$ upon lowering the temperature.  The interactions are treated in a generalized random-phase 
approximation (see Ref. \cite{Bulut:2013} and \ref{sec:A2}), which allows one to find
the leading charge instability without any bias towards a particular ordering wavevector 
${\bf q}^*$ or orbital type. 

\section{Results}
\label{sec:results}

The main results of this calculation are summarized in Fig.~\ref{fig:pdep}. The Fermi 
surface for the ALJP bands is shown in Fig.~\ref{fig:pdep}(a), along with the wavevectors 
$\bq_1$ and $\bq_2$ at which the charge susceptibility first diverges upon cooling in the 
absence of staggered Cu moments. As in previous calculations \cite{Metlitski:2010vf,Efetov:2013,Meier:2013,Holder:2012ks,Husemann:2012vg,Sachdev:2013,Bulut:2013,Sau:2013vw}, these 
wavevectors lie along the BZ diagonals and the charge instability primarily involves an 
intra-unit cell charge transfer between O$p_x$ and O$p_y$ orbitals. ${\bf q}_1$ and 
${\bf q}_2$ connect points close to nearby hot-spot regions of the Fermi surface.
When $M$ is finite but small, as in Fig.~\ref{fig:pdep}(b), the Fermi surface breaks up into
hole pockets around $(\pm\pi/2,\pm\pi/2)$ and electron pockets centered at the ``antinodal'' 
points on the BZ boundary; the modulation wavevectors remain diagonal and connect these 
pockets.

While the directions of $\bq_1$ and $\bq_2$  are consistent with previous calculations,
they conflict with experiments \cite{Ghiringhelli:2012bw,Chang:2012vf,Achkar:2012ep,Wu:2013,Blanco-Canosa:2013,Blackburn:2013}, which clearly indicate that the charge densities 
are modulated along the {\em axial} Cu-O bond directions. This discrepancy is resolved when 
the electron pockets are fully eliminated [Fig.~\ref{fig:pdep}(c)] by a sufficiently large 
staggered potential $M$ and the modulation wavevectors rotate to the axial direction.
Furthermore, the magnitude of $\bq_{1,2}^\ast$ agrees quantitatively with the experimental 
data of Blackburn {\em et al.} \cite{Blackburn:2013} as shown in Fig.~\ref{fig:pdep}(d) for 
the doping dependence of $|\bq_{1,2}^\ast|$. We emphasize that no fine tuning of the model 
parameters was done to obtain these results: the band parameters were taken from 
Ref.~\cite{Andersen:1995}, and $\bq_1^\ast$ and $\bq_2^\ast$ depend only weakly on the size of 
$M$ once it is large enough to remove the electron pockets.
 
 The fact that the Fermi surface forms well defined pockets is an artefact of the assumption that the AF correlation length $\xi_\mathrm{AF}$ is infinite\cite{Schmalian:1999}, and indeed there is no experimental evidence for hole pockets of the type shown in Fig.~\ref{fig:pdep}(c).  When the $\xi_\mathrm{AF}$ is finite, however, the pockets become arcs, which is consistent with experiments.  For our purposes, it is important to note that  the portions of the Fermi surface connected by $\bq_1^\ast$ and $\bq_2^\ast$ remain well defined when $\xi_\mathrm{AF}$ is finite\cite{Schmalian:1999}, while the back sides of the pockets are wiped out.  For this reason, we believe that the leading charge instability described here will also be the leading instability in models with short range AF correlations.
  
The charge modulation amplitudes on the different orbitals are determined from the 
eigenvector ${\bf v}_j^\chi$ of the divergent eigenvalue of the $3\times 3$ susceptibility 
matrix $\chi_{\alpha\beta}(\bq_j^\ast)$ ($j=1,2$) at the transition. The three components of 
${\bf v}_j^\chi$ give the {\em relative} (but not absolute) modulation amplitudes $\delta
n_\mathrm{Cu}(\bq)$, $\delta n_x(\bq)$, and $\delta n_y(\bq)$. A purely nematic mode, with 
$d$-wave charge transfer between O$p_x$ and O$p_y$ orbitals only and no modulation on the 
Cu$d$ orbitals, would have an eigenvector ${\bf v}^\chi = (0,-1,1)/ \sqrt{2}$. For 
comparison, the calculated eigenvectors are ${\bf v}_1^\chi= (0.21, -0.39, 0.89)$ and 
${\bf v}_2^\chi = (0.21, 0.89, -0.39)$ when $p=0.10$ and  $M=1.5$ eV.   Thus, for $\bq^\ast_1$, the charge modulation amplitudes on the Cu and O$p_x$ sites are about 25\% and 43\% respectively of the amplitude on the O$p_y$ site.  This distinction between O$p_x$ and O$p_y$ sites is  consistent with the observation of anisotropic NMR linewidths in YBa$_2$Cu$_3$O$_{6.5}$  \cite{Wu:2014vx}:  the linewidths of O(2) oxygen sites,
 which lie perpendicular to $\bq^\ast$\cite{Blackburn:2013}  are roughly 50\% greater than for O(3) sites, which lie parallel to $\bq^\ast$.  As pointed out in \cite{Thomson:2014}, the ratio of the O$p_y$ to O$p_x$ modulation grows (shrinks) rapidly with increasing (decreasing) $U_d$. 

Figure~\ref{fig:pdep}(e) illustrates the unidirectional charge modulations derived from 
${\bf v}^\chi_1$. As ${\bf v}_1^\chi$ directly tells, the charge modulations on the O$p_x$ and 
O$p_y$ orbitals are out of phase, so there is a significant intra-unit cell nematic-like 
charge transfer between them. The charge ordered phase is not purely nematic, however, as 
there are also modulations of the total charge per unit cell and of the Cu charge. This 
structure is consistent with the observation of nematic-like modulations of the oxygen 
orbitals by STM \cite{Kohsaka:2007hx,Wise:2008,Lawler:2010n} and elastic RXS 
\cite{Comin:2014vq}, and the observation of Cu charge modulations by NMR \cite{Wu:2011ke}. 
For a unit cell centered on a Cu$d$ orbital at $\brr$, the total charge modulation is 
$\delta n_\mathrm{tot}(\brr)=\delta n_\mathrm{Cu}(\brr)+\frac 12\sum_{\bdelta} \delta n_p({\brr +{\bdelta}})$, where ${\brr+ {\bdelta}}$ are the 
locations of the four neighboring oxygen atoms; the nematic modulation is defined by 
$\delta n_\mathrm{nem}(\brr)=\frac 12 \sum_{\bdelta}(-1)^{\delta_y} \delta n_p({\brr + 
{\bdelta}})$. Figure~\ref{fig:pdep}(f) clearly shows that all three types of
modulation are present. These different symmetries must in fact mix because 
$\chi_{\alpha\beta}(\bq)$ is not invariant under fourfold rotations when $\bq\neq {\bf 0}$.

\begin{figure}[tb]
\begin{center}
\includegraphics[width=0.5\columnwidth]{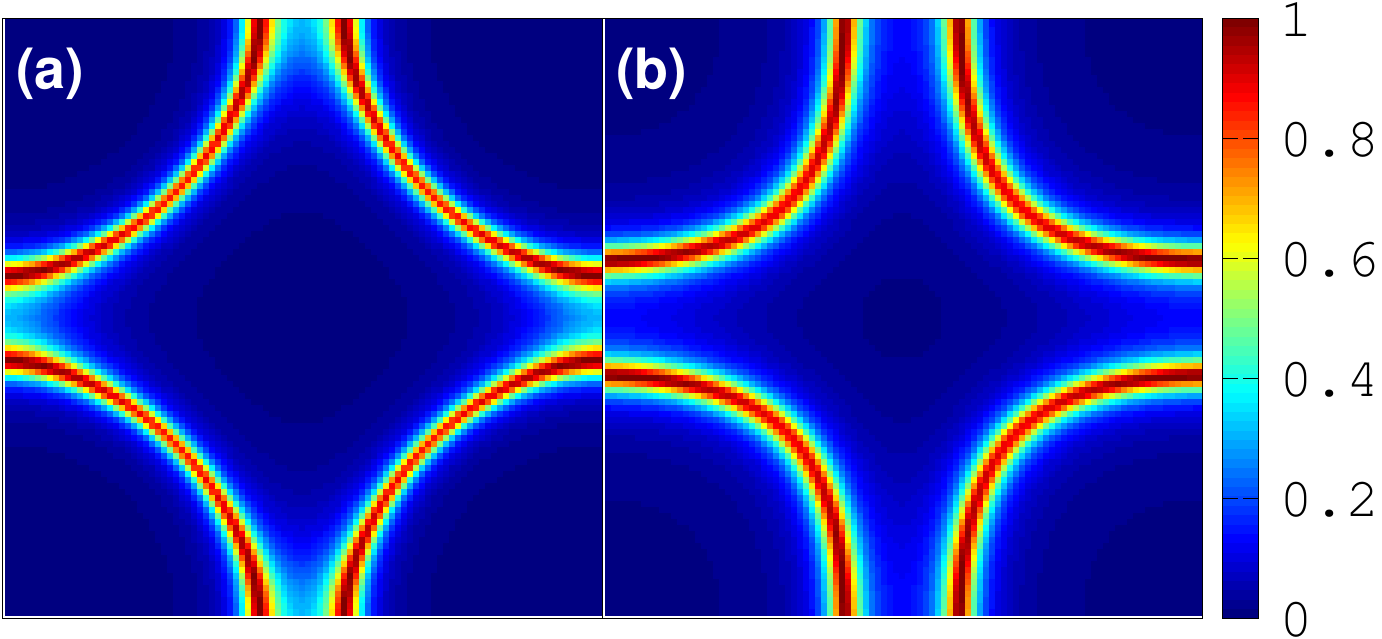}
\end{center}
\caption{Spectral functions $A_\mathrm{Cu}(\bk,\varepsilon_F)$ at the Fermi energy, projected 
  onto $\mathrm{Cu}$ orbitals. Results are for (a) the Emery model and (b) the ALJP model.  
  In both models, $\tilde t_{pp} = t_{pp}^d + t_{pp}^i = -1.0$ eV: in (a) $t_{pp}^d=-1.0$ eV 
  and $t_{pp}^i=0$; in (b)  $t_{pp}^d=0$ and $t_{pp}^i=-1.0$ eV. Other parameters are 
  $t_{pd}=1.6$ eV and $\epsilon_d-\epsilon_p=0.9$ eV.  }
\label{fig:Fermisurface}
\end{figure}

To understand the role of the Cu4s orbital, we compare our results to those for the Emery 
model, which does not include it. $\tilde t_{pp}=-1$~eV is chosen for both models, so that
the only difference between them is that the diagonal matrix elements of ${\bf H}(\bk)$ are 
unrenormalized in the Emery model. As shown in Fig.~\ref{fig:Fermisurface}, this 
changes the Fermi-surface shape and the underlying band structure only quantitatively, with 
a noticeable increase of the Fermi-surface curvature. Indeed, the incommensurate peak 
positions $\bq^\ast_j$ in the charge susceptibility shift only by about 5\% between the two 
models for $M=1.5$ eV. Surprising and important, however, is that the leading instability in
the Emery model is to a ${\bf q}={\bf 0}$ nematic phase, and that the 
incommensurate phase is subleading. We have traced this difference to the oxygen spectral 
weight distribution along the Fermi surface, which is strongly anisotropic in the Emery 
model, but nearly isotropic in the ALJP model (see \ref{sec:A1}). Thus, the
Cu4s orbital stabilizes the ALJP model against $\bq={\bf 0}$ nematic order.

\begin{figure}[tb]
\begin{center}
\includegraphics[width=0.5\columnwidth]{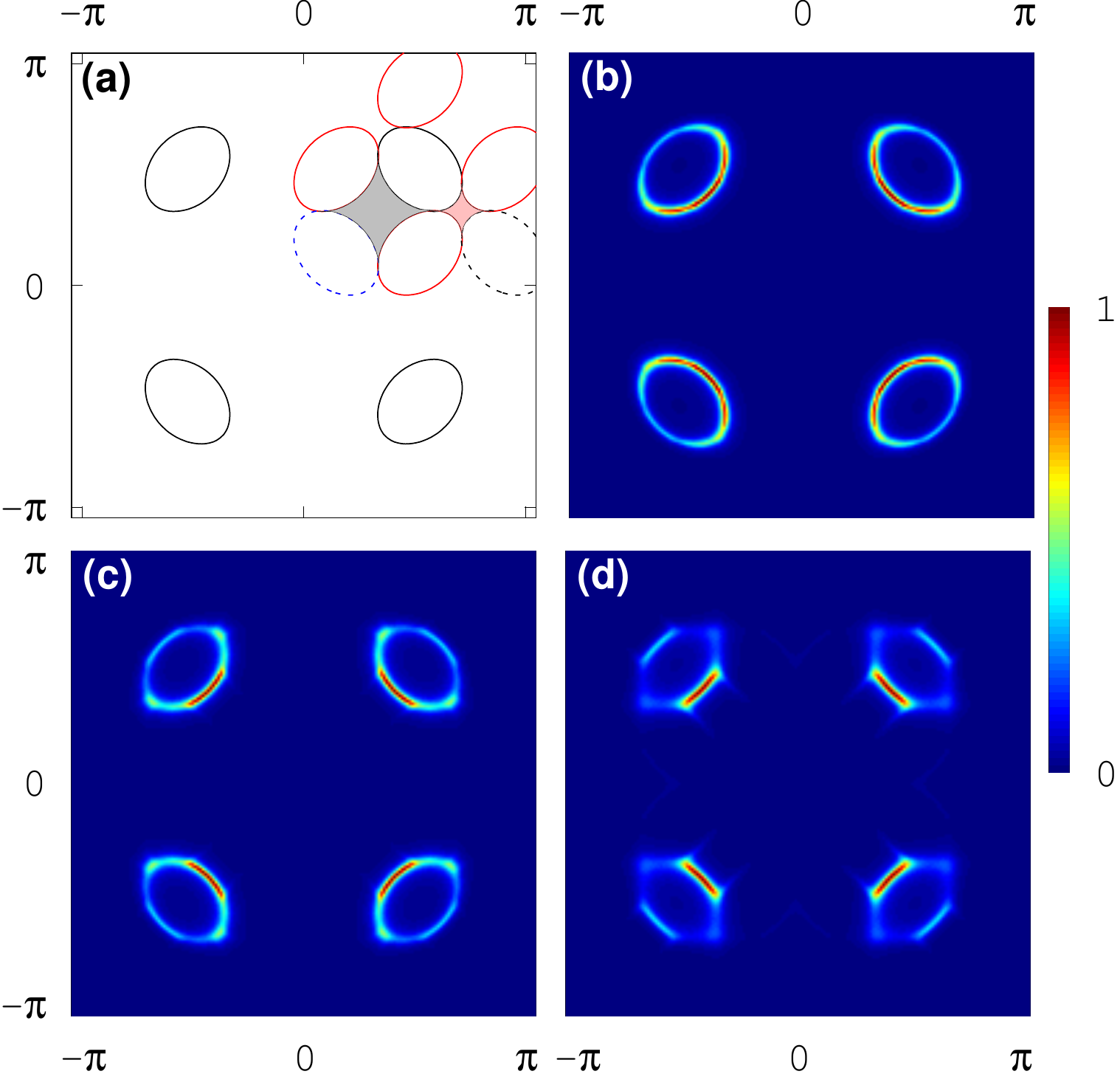}
\end{center}
\caption{Fermi surface and spectral function in the charge ordered state at $p=0.10$. (a) Nodal Fermi 
  pockets of the ALJP model with $M=1.5$ eV (black) along with some of the first- (red) and 
  second-order (dotted) replica Fermi surfaces that are involved in the reconstruction of 
  the $(1,1)$ nodal pocket by charge order. The first-order replicas shown are obtained by 
  shifting nodal pockets by $\pm\bq_1^\ast$ or $\pm \bq_2^\ast$. Second-order replicas result
  from shifting the $(-1,-1)$ pocket by $\bq_1^\ast + \bq_2^\ast$ (blue dotted) or 
  $-\bq_1^\ast+\bq_2^\ast$ (black dotted). These replicas bound electron pockets (shaded grey 
  and pink regions) with areas $A_1 = 0.50/a_0^2$ and $A_2 = 0.10/a_0^2$. (b)-(d) Spectral
  functions at the Fermi energy for bi-directional charge order with modulation potential 
  (b) $\delta\epsilon=0$, (c) $\delta\epsilon=0.25$ eV, and (d) $\delta \epsilon = 0.5$ eV.
  The spectral function is broadened by $0.04$ eV.}
\label{fig:specfun}
\end{figure}

To discuss the Fermi surface reconstruction from charge order we show in 
Fig.~\ref{fig:specfun}(a) the four original Fermi surface hole pockets centered at $(\pm
\pi/2,\pm \pi/2)$, which we label $(\pm 1,\pm 1)$; these are the ``nodal'' pockets. Charge 
order along a direction $\bq_j^\ast$ scatters quasiparticles through $\pm \bq_{j}^\ast$ and 
generates replica Fermi-surface pockets. Red contours mark those first-order replicas, 
generated by shifting the $(-1,1)$ pocket by $\pm\bq^\ast_1$ and the $(1,-1)$ pocket by 
$\pm\bq^\ast_2$, that touch the $(1,1)$ nodal pocket. Where original and replica pockets 
touch, the bands hybridize and a gap opens. Importantly, at any doping $\bq_{1,2}^\ast$ are 
such that replica and original pockets precisely touch without crossing. We include also a 
second-order replica (blue dotted) by shifting the $(-1,-1)$ pocket by 
$\bq^\ast_1 +\bq^\ast_2$. This replica appears only when the order is bi-directional, 
and it hybridizes with two of the first order replicas and the original $(1,1)$ nodal hole 
pocket to form a diamond-shaped electron pocket shown as the grey region on the {\em front} 
side of the $(1,1)$ pocket [closest to the origin] in Fig.~\ref{fig:specfun}(a).

It was argued empirically \cite{Sebastian:2012wh} that electron pockets of this diamond type
could explain observed magneto-oscillations in YBa$_2$Cu$_3$O$_{6.5}$. Yet, the interpretation
is complicated because, in addition to a central frequency of $F_\mathrm{expt}\sim 530$ T
\cite{DoironLeyraud:2007bj,Singleton:2010bz,Riggs:2011ii}, a pair of side frequencies 
is observed \cite{Audouard:2009df}. The latter have been attributed to bilayer splitting 
into bonding and antibonding bands \cite{Audouard:2009df,Sebastian:2012wh}. For the ALJP 
model, we find that the electron pocket has an area $A_1 =0.50/a_0^2$ ($a_0$ is the lattice 
constant) which gives an oscillation frequency $F_1 = (\hbar/2\pi e)A_1 = 340$ T, slightly 
less than $F_\mathrm{expt}$. However, since $A_1$ represents only $\sim 1\%$ of the BZ area, 
it is far more sensitive to the Fermi surface shape than is $\bq^\ast$. We obtain, for 
example, $F_1 = 1000$ T using the Emery model with $M=1.5$eV and the parameters in the 
caption of Fig.~\ref{fig:Fermisurface}; this is a factor of 3 larger than the ALJP result, 
even though the incommensurate $q^\ast$ differs by only $\sim 10\%$ between the two models. 
Obviously, fine tuning of the ALJP model, which is based on band structure calculations for 
YBa$_2$Cu$_3$O$_7$, is needed to quantitatively match quantum oscillation experiments 
performed on YBa$_2$Cu$_3$O$_{6.5}$. 
 
One difference to the proposal in Ref.~\cite{Sebastian:2012wh} is that we find four electron
pockets attached to each nodal pocket, rather than one. In addition to the electron pocket 
discussed above, there is a second electron pocket with identical area (not shown) on the 
{\em back} side of the nodal pocket [closest to $(\pi,\pi)$]. Two further diamond-shaped
electron pockets form at opposite ends of the each nodal pocket. One of these, with an area
area $A_2 = 0.10/a_0^2$ and corresponding oscillation frequency $F_2 = 65$ T, is shown as 
a shaded pink region in Fig.~\ref{fig:specfun}(a). These additional electron pockets are
an artefact of the assumed infinite AF correlation length. As we said previously, when $\xi_\mathrm{AF}$ is 
finite, the spectral function is characterized by Fermi arcs that resemble the front side of
the nodal pockets; the back and side electron pockets only emerge as $\xi_\mathrm{AF}$ 
diverges.\cite{Schmalian:1999}  
 
To see the effect of charge order on the spectral function, we model bi-directional charge 
order as a perturbation of the Cu$d$, O$p_x$, and O$p_y$ site energies by 
$\delta\epsilon [{\bf v}_1^\chi\cos(\bq_1^\ast\cdot \brr)+{\bf v}_2^\chi
\cos(\bq_2^\ast\cdot\brr)]$. Adding the corresponding potential term to the Hamiltonian, we
calculate the spectral function $A(\bk,\omega)=\sum_\alpha\sum_n|\phi_{\alpha n}(\bk)|^2 
\delta(\omega-E_{n\bk})$ at the Fermi energy $\omega=\varepsilon_F$, where $\phi_{\alpha n}
(\bk)$ are the energy eigenvectors indicating the projection of band $n$ onto orbital 
$\alpha$, and $E_{n\bk}$ are the energy eigenvalues. Figure~\ref{fig:specfun}(b) shows 
$A(\bk,\varepsilon_F)$ without charge order ($\delta\epsilon=0$).
In Figs.~\ref{fig:specfun}(c) and (d) the modulation potential is increased to 
$\delta\epsilon = 0.25$ eV and $\delta \epsilon=0.5$ eV, respectively. These selected values
are exaggerated for presentation purposes. The main effect of charge order is to erode
spectral weight along segments of the Fermi surface that touch replicas as in 
Fig.~\ref{fig:specfun}(a). In contrast, the spectral weight is almost unaffected by charge 
order along short arcs on the insides of the nodal pockets. Also, the diamond-shaped
electron pockets shown in Fig.~\ref{fig:specfun}(a) are unobservable, even for the 
unphysically large value of $\delta \epsilon$ used in Fig.~\ref{fig:specfun}(d).
  
\begin{figure}
\begin{center}
\includegraphics[width=0.4\columnwidth]{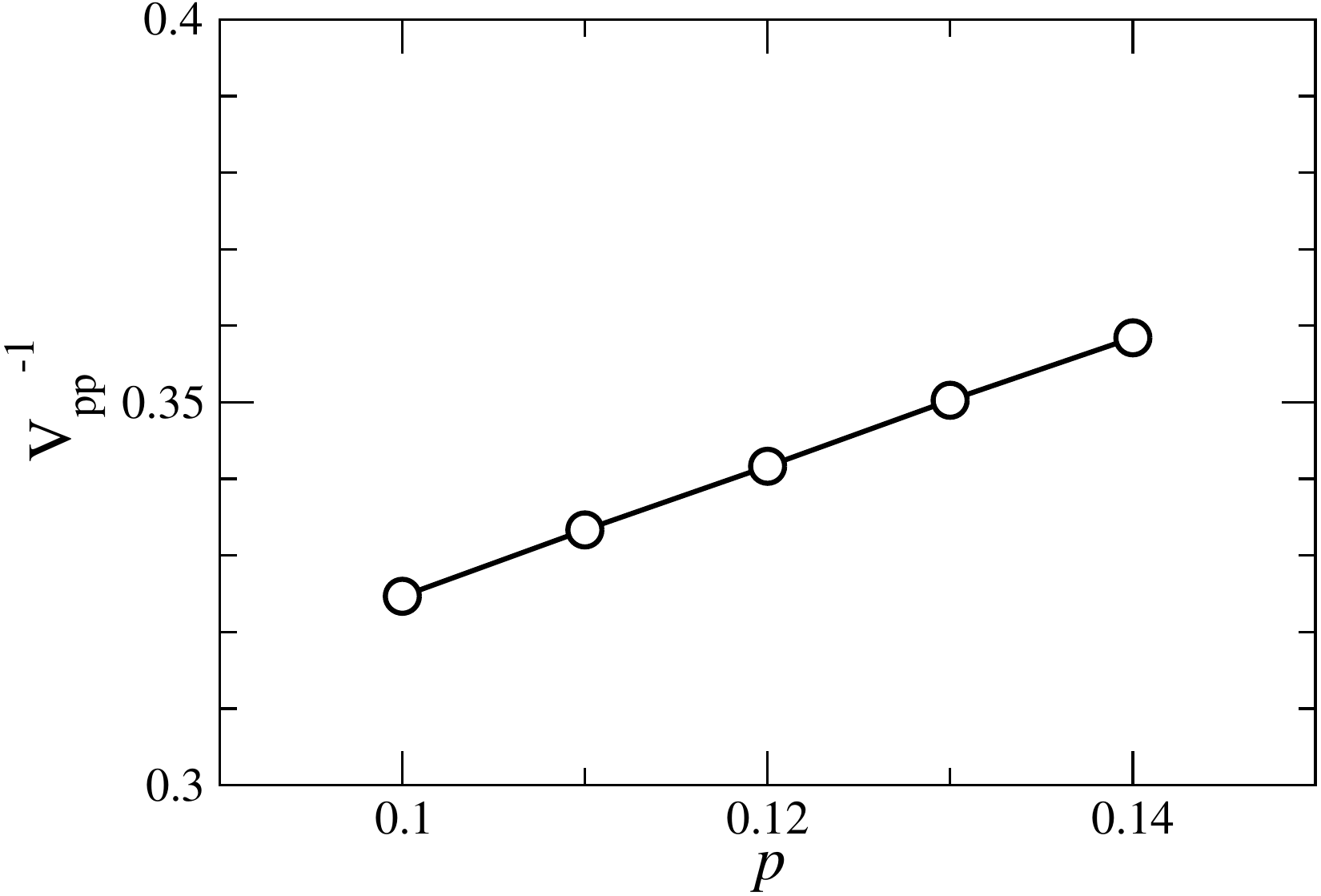}
\end{center}
\caption{Doping dependence of the critical value of $V_{pp}^{-1}$ in units of [eV]$^{-1}$ at 
$T=110$~K for $M=1.5$~eV.}
\label{fig:Vpp_vs_p}
\end{figure}

In our model calculations, the charge instability is driven by the Coulomb repulsion $V_{pp}$ 
between electrons on neighboring oxygen atoms. In the doping window $0.1<p<0.14$ the ordering 
wavevector $\bq_j^\ast$ continuously decreases with $p$ as in the x-ray diffraction experiments 
by Blackburn {\em et al.} \cite{Blackburn:2013} [Fig. 1(d)]. In the same doping regime the 
calculated charge ordering temperature $T_\mathrm{co}$ rises with increasing $p$. Because the 
calculation of $T_\mathrm{co}$ is numerically intensive, we show instead in 
Fig.~\ref{fig:Vpp_vs_p} the inverse of the critical interaction strength, namely $V_{pp}^{-1}$, 
required to drive the charge ordering transition at fixed $T=110$~K. This quantity is a useful
proxy for $T_\mathrm{co}$: a large value of $V_{pp}^{-1}$ indicates that the system is very 
susceptible for charge ordering, and should therefore have a large $T_\mathrm{co}$. In our 
calculations, the susceptibility towards charge order with growing hole density $p$ 
concomitantly increases with the increasing size of the nodal hole-Fermi pockets.

Experimentally, the variation of $T_\mathrm{co}$ with hole doping remains inconclusive. RXS 
data indicate that $T_\mathrm{co}$ decreases with increasing $p$ \cite{CominScience2014}, but 
this trend is at variance with earlier x-ray data and with the field-tuned $T_\mathrm{co}(H)$
observed by NMR \cite{Wu:2011ke,Wu:2013}. From the latter data a maximum $T_\mathrm{co}$ around 
$p=0.12$ was inferred \cite{Wu:2013}, and a similar dome-shaped $p$-dependence was determined 
for the Fermi-surface reconstruction from Hall measurements \cite{Wu:2011ke}. Recent x-ray 
experiments on YBa$_2$Cu$_3$O$_{6+x}$ \cite{Huecker:2014vc} also find a dome-shaped dependence of $T_\mathrm{co}$ on $p$, peaked at $p \sim 0.10$.
 
The evolution of $T_\mathrm{co}$  in model calculations likely depends on the detailed doping dependence of both the 
effective interaction strength in the charge ordering channel,  which has not been considered here, and the Fermi surface,  which is the central topic of this work. A further 
complication is the role of disorder, which is unavoidable in doped cuprates and should 
influence the spatial lock-in of any charge-density wave. The issue of how $T_\mathrm{co}$ evolves with doping is an open question that needs to be resolved.

\section{Conclusions}
\label{sec:conclusions} 
In this work, we have described a model calculation that provides a route to understand the 
doping dependence of the charge-ordering wavevectors $\bq^\ast$ in cuprate superconductors. The 
essential model ingredients are a realistic multiorbital description of the CuO$_2$ planes, 
the assumption and the ansatz that strong correlation effects on the Cu$d_{x^2-y^2}$ orbitals 
can be modeled by antiferromagnetically correlated moments, and the inclusion of short range 
Coulomb forces that drive the charge-ordering instability. While the model analysis is still 
incomplete, e.g. inelastic spin-scattering processes and the spin dynamics are neglected, it 
nonetheless provides an important result: quantitatively correct charge-ordering wavevectors 
$\bq^\ast$ are obtained, if the charge order is presumed to emerge {\em from} the pseudogap 
phase, rather than to generate the pseudogap itself.

Also a subtle but important role played by multiorbital physics is highlighted. While the 
three-orbital Emery model and the four-orbital ALJP model have similar Fermi surfaces, the 
leading instability in the Emery model is to a $\bq=0$ nematic phase, while the ALJP model 
correctly reproduces the structure seen experimentally. This distinction is traced to subtle 
differences in the orbital composition of the conduction band.

A number of questions necessarily remains open, in particular the relationship between the 
charge order and the pseudogap and also the possible connection to the emergence of 
spontaneous loop currents await further clarification. Notably, the dependence of 
$T_\mathrm{co}$ on $p$ is a challenging question that demands an improved treatment of the 
pseudogap phase beyond the initial steps presented in this work.  

\section*{Acknowledgments}
We acknowledge helpful conversations with A. V. Chubukov, M.-H.\ Julien, C. P\'epin, A.\ Thomson and S. Sachdev.   We particularly thank A.\ Thomson and S.\ Sachdev for pointing out an error in the original calculation.
W.A.A. acknowledges support by the Natural Sciences and Engineering Research Council (NSERC)
of Canada.  A.P.K. acknowledges support by the 
Deutsche Forschungsgemeinschaft through TRR 80.

\appendix
\setcounter{section}{0}
\section{Model and Band Structure}
\label{sec:A1}
\subsection{Effective Three-Band Model}
\begin{figure}[tb]
\includegraphics[width=0.5\columnwidth]{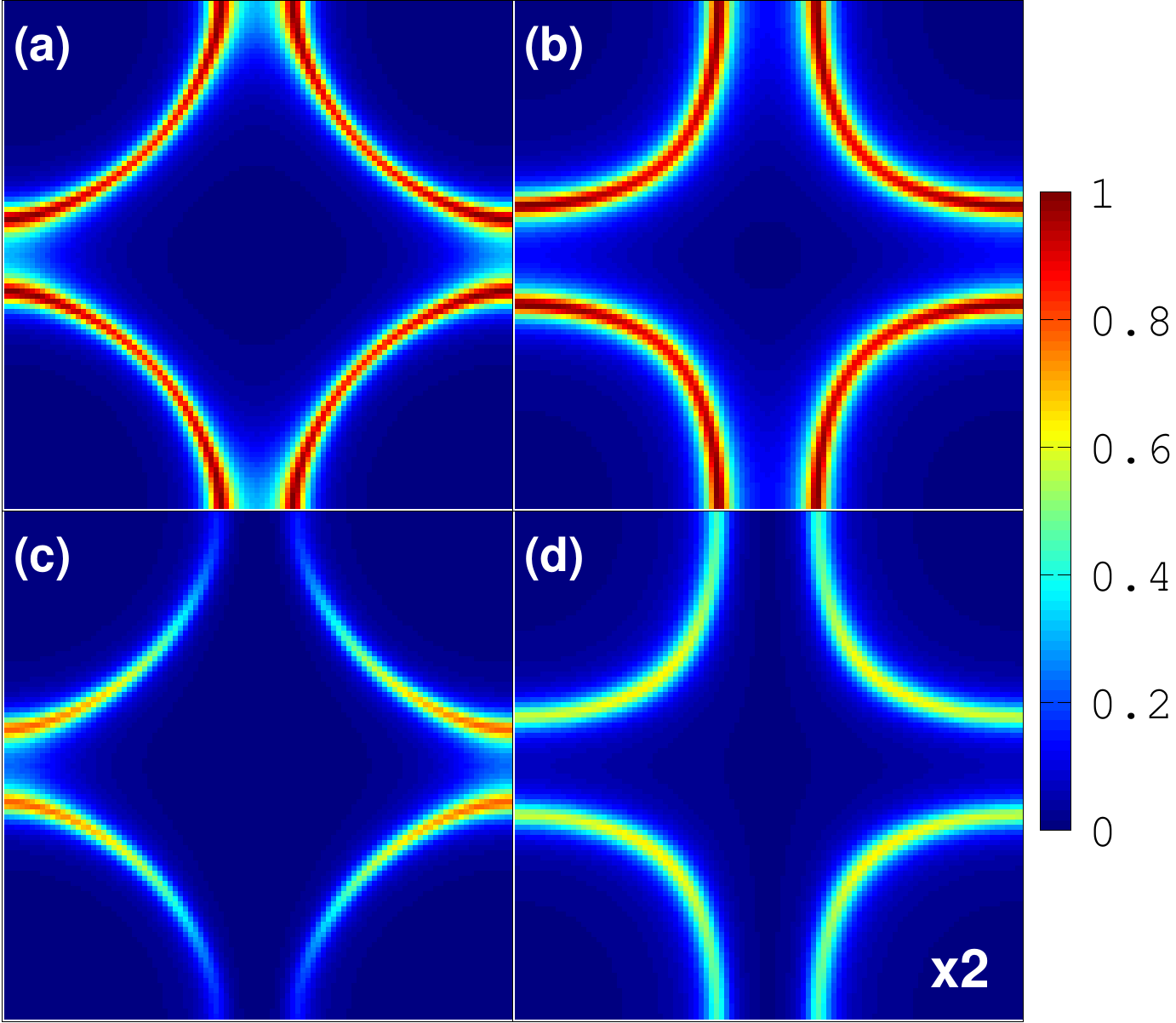}
\caption{Spectral functions (a),(b) $A_\mathrm{Cu}(\bk,\varepsilon_F)$ and
(c),(d) $A_\mathrm{p_x}(\bk,\varepsilon_F)$
 at the Fermi energy, projected 
  onto $\mathrm{Cu}$  and O$p_x$ orbitals respectively. 
  Results are for (a),(c) the Emery model and (b),(d) the ALJP model.  
  As in the main text,  $\tilde t_{pp} = t_{pp}^d + t_{pp}^i = -1.0$ eV in both models. }
\label{fig:expandedFS}
\end{figure}

We start with a realistic four-band model that is tailored specifically to YBa$_2$Cu$_3$O$_7$, due to 
Andersen, Liechtenstein, Jepsen, and Paulsen\cite{Andersen:1995} (ALJP).  In addition
to the copper $3d_{x^2-y^2}$ and oxygen $p_x$ and $p_y$ orbitals included in the usual 
three-band Emery model \cite{Emery:1987prl}, the ALJP model includes the Cu$4s$ orbital.  
The $4s$ orbital lies $\sim 6.5$ eV above the $d_{x^2-y^2}$ orbital and is often ignored; however,
band structure calculations\cite{Andersen:1995} showed that indirect hopping through
the $4s$ orbital between neighboring O$p_x$ and O$p_y$ orbitals is actually larger than
the direct hopping.  The four-band Hamiltonian is 
\begin{equation}
{\bf \hat H_\mathrm{4b}} = \sum_\bk \tilde \Psi^\dagger_{\bk} \left [
\begin{array}{cccc} 
	\epsilon_{d} & 2t_{pd} s_x & -2t_{pd} s_y & 0 \\
	2t_{pd} s_x & \epsilon_{x} &   4t_{pp} s_x s_y & 2 t_{ps} s_x \\
	-2t_{pd} s_y & 4t_{pp} s_x s_y & \epsilon_{y} & 2t_{ps} s_y \\
        0 & 2 t_{ps} s_x & 2t_{ps} s_y & \epsilon_s
\end{array}
\right ]  \tilde \Psi_\bk,
\label{H4b}
\end{equation}
where $s_x =\sin (k_x/2)$ and $s_y=\sin(k_y/2)$, and where
\begin{equation}
\tilde \Psi_{\bk} = \left [\begin{array}{c} d_{\bk} \\ p_{x\bk} \\ p_{y\bk} \\ s_{\bk}
\end{array} \right ],
\label{Psitilde}
\end{equation}
is an array of electron annihilation operators for the Cu$3d_{x^2-y^2}$, O$p_x$, O$p_y$, and
Cu$4s$ orbitals, respectively. The spin index is suppressed in Eqs.~(\ref{H4b}) and 
(\ref{Psitilde}).

We can integrate out the $4s$ orbital in the usual downfolding procedure 
\cite{Andersen:1995,Loewdin:1951}. Writing the four-band Hamiltonian matrix in a block 
structure,
\begin{equation}
\left [ \begin{array}{cc}
{\bf H^0}(\bk) _{3\times 3} & {\bf  H^\perp}(\bk) _{3\times 1} \\
{{\bf H^\perp}(\bk) }^\dagger_{1\times 3} & \epsilon_s 
\end{array} \right ]
\end{equation}
where the subscript notation $i\times j$ denotes the size of each block, we solve the 
equations-of-motion for the Green's function in the subspace of $d_{x^2-y^2}$, $p_x$, 
and $p_y$ orbitals:
\begin{equation}
{\bf G}(\bk,\omega) = \left [ \omega {\bf 1} -{\bf H^0}(\bk) 
- {\bf H^\perp}(\bk) \frac{1}{\omega-\epsilon_s} {{\bf H^\perp}(\bk)}^\dagger \right ]^{-1}_{3\times 3}.
\end{equation}
From the structure of ${\bf G}(\bk,\omega)$ at $\omega=\varepsilon_F$, 
an effective three-band Hamiltonian matrix is generated
\begin{eqnarray}
{\bf H}(\bk) = 
{\bf H^0}(\bk) 
+ {\bf H^\perp}(\bk) \frac{1}{\omega-\epsilon_s} {{\bf H^\perp}(\bk)}^\dagger \nonumber \\[1mm]
= \left[ \begin{array}{ccc} 
    \epsilon_{d} & 2t_{pd} s_x & -2t_{pd} s_y \\
    2t_{pd} s_x & \tilde \epsilon_{x}(\bk) &   4\tilde t_{pp} s_x s_y \\
    -2t_{pd} s_y & 4\tilde t_{pp} s_x s_y & \tilde \epsilon_{y}(\bk)
\end{array}
\right ]
\label{eq:H}
\end{eqnarray}
with 
\begin{eqnarray}
\tilde \epsilon_x(\bk) &=& \epsilon_p + 4t_{pp}^i s_x^2 ,\label{eq:ex} \\
\tilde \epsilon_y(\bk) &=& \epsilon_p + 4t_{pp}^i s_y^2, \label{eq:ey} \\
\tilde t_{pp} &=& t_{pp}^d + t_{pp}^i, 
\end{eqnarray}
where $t_{pp}^d$ is the direct hopping between $p_x$ and $p_y$ orbitals, and
\begin{equation}
t_{pp}^i = \frac{t_{ps}^2}{\varepsilon_F-\epsilon_s}
\end{equation}
is the indirect hopping amplitude, through the $4s$ orbital, between $p$ orbitals.
Importantly, we note that
$\varepsilon_F < \epsilon_s$, so that 
\begin{equation}
t_{pp}^i  < 0.
\end{equation}
Based on the signs of the orbital lobes, we would expect $t_{pp}^d > 0$; however, Andersen 
{\em et al.} proposed that $t_{pp}^d$ is negligible compared to the indirect contribution, 
and that $\tilde t_{pp}\sim -1$ eV.   Throughout this work, we adopt the values
of $t_{pd}$, $t^d_{pp}$, $t^i_{pp}$, and $\epsilon_d-\epsilon_p$ 
given by ALJP\cite{Andersen:1995} and listed in Table~\ref{table:param}.
Figure~\ref{fig:expandedFS} shows the spectral functions at the Fermi energies projected
onto both Cu and O${p_x}$ orbitals.  For comparison, results are also shown for the Emery
model.

\begin{table}
\begin{tabular}{l|l}
Parameter & Value (eV) \\
\hline
 $t_{pd}$ & 1.6 \\
 $t_{pp}^d$ & 0  \\
$t_{pp}^i$ &-1.0  \\
 $\epsilon_d-\epsilon_p$ & 0.9  \\
 $M$ & 0.0-1.5 \\
 $U_d$ & 6.0\\
 $U_p$ & 3.0 \\
 $V_{pd}$ & 1.0 \\
 $V_{pp}$ & variable \\
 \hline
\end{tabular}
\caption{Model parameters used in this work.}
\label{table:param}
\end{table}

\subsection{Slater Antiferromagnetism}
\begin{figure}
\includegraphics[width=0.5\columnwidth]{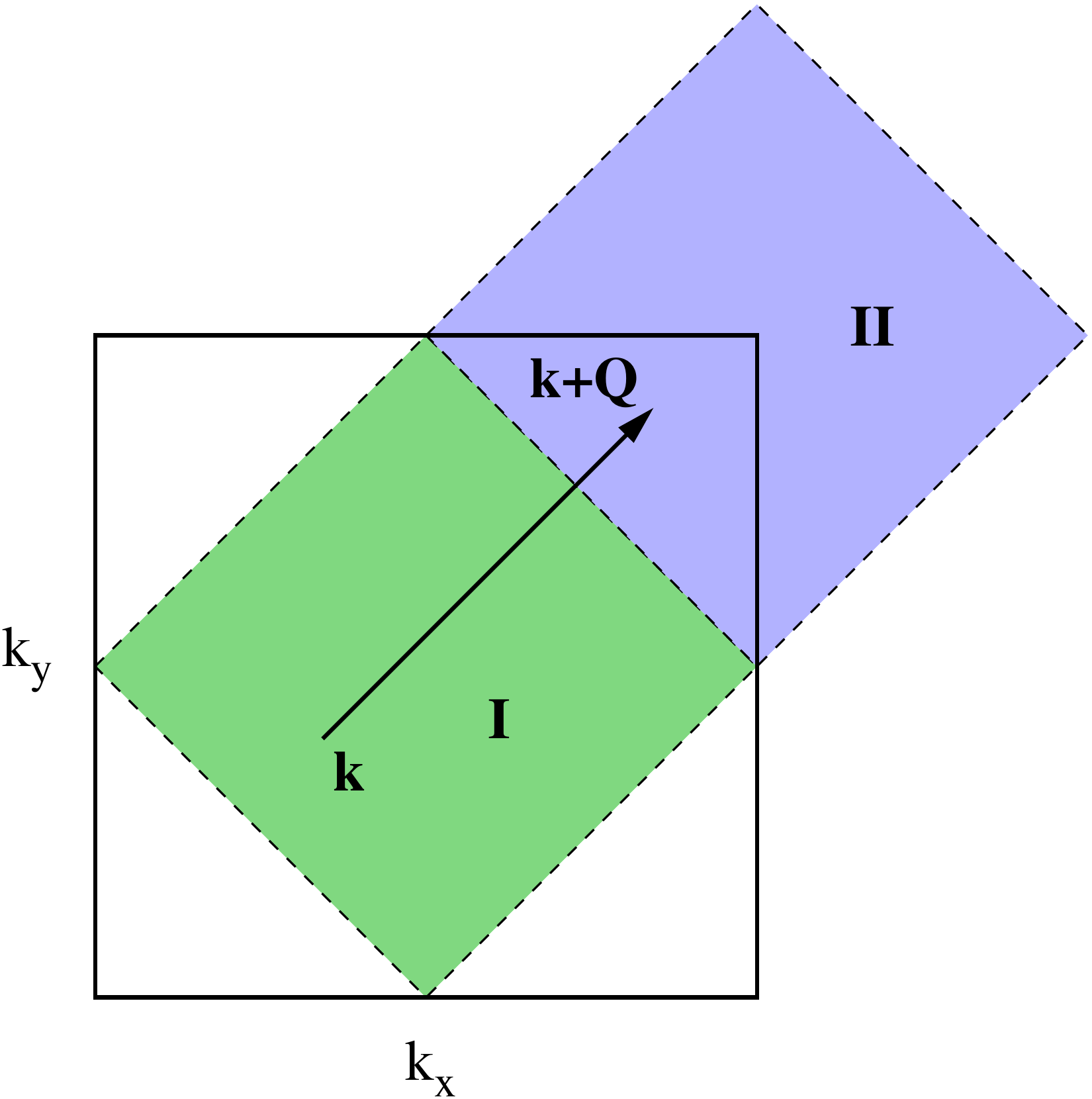}
\caption{Reduced AF BZ. The black square shows the original BZ for the nonmagnetic lattice.  
The areas labelled I and II are the first and second AF BZs.}
\label{fig:AFBZ}
\end{figure}
We add a staggered magnetic field at the copper sites to the Hamiltonian to generate local 
moments on the Cu$d$ orbitals. Then, with the spin index included, the Hamiltonian is
\begin{equation}
\hat H_\mathrm{6b} = \sum_{\bk,\sigma} \tilde \Psi^\dagger_{\bk \sigma}
\left [\begin{array}{cc} 
{\bf H}(\bk) & -\sigma {\bf M} \\ 
-\sigma {\bf M} & {\bf H}(\bk+\bQ) 
\end{array}\right ] 
\tilde \Psi_{\bk \sigma}
\label{eq:6b}
\end{equation}
where $\bQ=(\pi,\pi)$, 
\begin{equation}
\tilde \Psi_{\bk\sigma} = 
\left [\begin{array}{c} \Psi_{\bk\sigma}  \\ \Psi_{\bk+\bQ\, \sigma} \end{array} \right ],
\label{eq:psitilde}
\end{equation}
and
\begin{equation}
{\bf M} = \left [\begin{array}{ccc} 
M & 0 & 0 \\
0 & 0 & 0 \\
0 & 0 & 0 
\end{array} \right ].
\end{equation}
In the state with staggered copper moments, the wavevector $\bk$ is restricted to the 
{\em first} antiferromagnetic (AF) BZ, labelled I in Fig.~\ref{fig:AFBZ}. Hence, 
$\bk+\bQ$ belongs to the {\em second} AF BZ, labelled II in Fig.~\ref{fig:AFBZ}.

\section{Generalized RPA}
\label{sec:A2}
\subsection{Diagrammatic perturbation theory}

\begin{figure*}[tb]
\includegraphics[width=\textwidth]{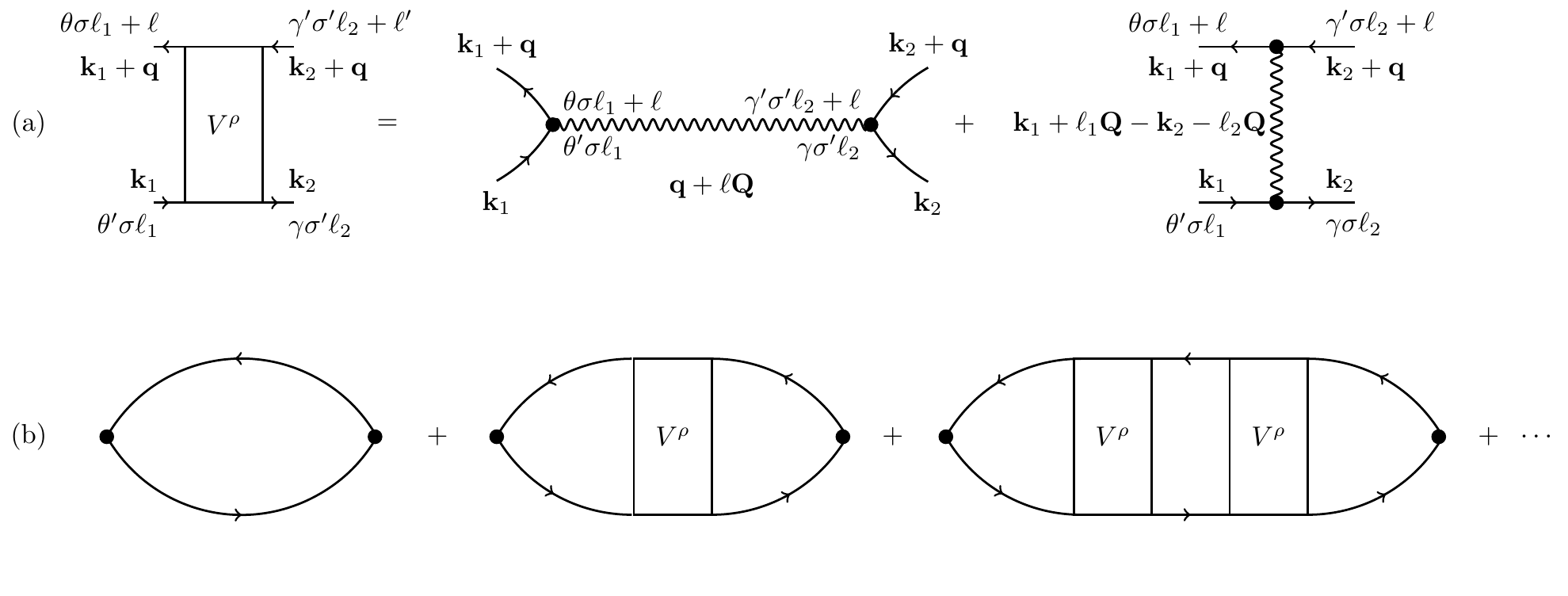}
\caption{ Diagrams evaluated in the calculation of the charge susceptibility.  
  (a) Effective interaction in the charge channel, including both Hartree (first term) 
  and exchange (second term) contributions. The wavevector $\bk$ is
  restricted to the first AFBZ, and the greek labels, denote the orbital type ($d$, $p_x$,
  $p_y$).  The labels $\ell_1$, etc.\, indicate the AFBZ of the corresponding electron creation or annihilation operator, which has momentum $\bk + \ell_1\bQ$.
    (b) Diagrams summed in the calculation of the charge susceptibility
  $\chi_{\alpha\beta}(\bq)$.}
\label{fig:chiRPA}
\end{figure*}

We calculate the nematic susceptibility by summing the ladder and bubble diagrams shown 
in Fig.~\ref{fig:chiRPA}. This is analogous to what was done in Ref.~\cite{Bulut:2013}, 
and we describe here how that calculation has been extended to the AF case.
 
In Fig.~\ref{fig:chiRPA}, the wavevectors $\bk$ and ${\bk + \bQ}$ are constrained to the 
first and second AFBZs, respectively, pictured in Fig.~\ref{fig:AFBZ}, while $\bq$ is 
unconstrained. 
In this notation $\bk$ is conserved along each propagator and the indices $\ell_1, \ell_2,\ldots=0,1$ label the AFBZ to which the creation or annihilation operators at the ends of the lines belong.  For example, the line end labeled $\bk_1\theta^\prime \ell_1$ in Fig.~\ref{fig:chiRPA} has a corresponding annihilation operator $c_{\theta^\prime \bk_1+\ell_1\bQ \sigma}$, where $\sigma$ is the electron spin.  
 
Fig.~\ref{fig:chiRPA}(a) shows the bare interaction vertex $V^\rho_{}(\bk,\bk^\prime,\bq)$ between 
charges, which includes both direct (first term) and exchange (second term) diagrams, which is
 \begin{eqnarray}
\fl [{V^\rho}]^{ \ell\sigma, \ell^\prime\sigma^\prime}_{\theta \gamma^\prime  \gamma  \theta^\prime}(\bk_1 \ell_1,\bk_2 \ell_2,\bq ) 
=
\delta_{\theta,\theta^\prime} \delta_{\gamma,\gamma^\prime}  \delta_{\ell,\ell^\prime}  V_{\theta\sigma, \gamma\sigma^\prime} (\bq+\ell\bQ) \nonumber \\
- 
\delta_{\ell,\ell^\prime} \delta_{\sigma,\sigma^\prime} 
 \delta_{\theta,\gamma^\prime} \delta_{\theta^\prime,\gamma} 
 V_{\theta\sigma, \gamma\sigma } (\bk_1 +\ell_1\bQ - \bk_2 - \ell_2\bQ)
 \label{eq:Vq}
\end{eqnarray}
where the first and second terms are the direct and exchange terms, respectively.  

In ${\bf q}$-space, the Coulomb interaction for the three-band model is
\begin{eqnarray}
\fl V_{\alpha\sigma, \beta\sigma^\prime}(\bq) = \left \{\begin{array}{ll}
U_d, & \alpha=\beta=d,\, \sigma = -\sigma^\prime \\
U_p, & \alpha=\beta=x,y,\, \sigma = -\sigma^\prime \\
2V_{pd}\cos(q_x/2), &  \alpha=x,\beta=d \mbox{ or } \alpha=d,\beta=x \\
2V_{pd}\cos(q_y/2), &  \alpha=y,\beta=d \mbox{ or } \alpha=d,\beta=y \\
4V_{pp}\cos(q_x/2)\cos(q_y/2), & \alpha=x,\beta=y\mbox{ or } \alpha=y,\beta=x
\end{array} \right . .
\end{eqnarray}

As in Ref.~\cite{Bulut:2013}, the sum in Fig.~\ref{fig:chiRPA}(b) is most easily
done by expressing the exchange and direct interactions in terms of a set of basis functions $g^i_{\alpha\beta}(\bk)$:
 \begin{eqnarray}
\fl   [{V^\rho}]^{ \ell\sigma, \ell^\prime \sigma^\prime}_{\theta \gamma^\prime \gamma \theta^\prime} (\bk_1 \ell_1 ,\bk_2 \ell_2,\bq) = 
 \delta_{\ell,\ell^\prime} \sum_{i,j=1}^{19}
 g^i_{\theta\theta^\prime}(\bk_1 + \ell_1\bQ) \tilde V^{i{\ell}\sigma, j{\ell}\sigma^\prime}(\bq+\ell\bQ) g^j_{\gamma^\prime \gamma}(\bk_2 + \ell_2\bQ),
\label{eq:exchange}
 \end{eqnarray}
 where 
\begin{eqnarray*}
g^1_{\alpha\beta}(\bk) = g^{12}_{\beta\alpha}(\bk) = \delta_{\alpha,d}\delta_{\beta,x} \cos(k_x/2)\\
 g^2_{\alpha\beta}(\bk) = g^{13}_{\beta\alpha}(\bk) = \delta_{\alpha,d}\delta_{\beta,x} \sin(k_x/2)\\
 g^3_{\alpha\beta}(\bk) = g^{14}_{\beta\alpha}(\bk) = \delta_{\alpha,d}\delta_{\beta,y} \cos(k_y/2)\\
 g^4_{\alpha\beta}(\bk) = g^{15}_{\beta\alpha}(\bk) = \delta_{\alpha,d}\delta_{\beta,y} \sin(k_y/2)\\
 g^5_{\alpha\beta}(\bk) = g^{16}_{\beta\alpha}(\bk) = \delta_{\alpha,x}\delta_{\beta,y} \cos(k_x/2)\cos(k_y/2)\\
 g^6_{\alpha\beta}(\bk) = g^{17}_{\beta\alpha}(\bk) = \delta_{\alpha,x}\delta_{\beta,y} \cos(k_x/2)\sin(k_y/2)\\
 g^7_{\alpha\beta}(\bk) = g^{18}_{\beta\alpha}(\bk) = \delta_{\alpha,x}\delta_{\beta,y} \sin(k_x/2) \cos(k_y/2)\\
 g^8_{\alpha\beta}(\bk) = g^{19}_{\beta\alpha}(\bk) = \delta_{\alpha,x}\delta_{\beta,y} \sin(k_x/2)\sin(k_y/2) \\
 g^9_{\alpha\beta}(\bk) = \delta_{\alpha,d} \delta_{\beta,d} \\
 g^{10}_{\alpha\beta}(\bk) = \delta_{\alpha,x} \delta_{\beta,x} \\
 g^{11}_{\alpha\beta}(\bk) = \delta_{\alpha,y} \delta_{\beta,y} 
\end{eqnarray*}

In this basis, the sum of the diagrams in fig.~\ref{fig:chiRPA}(b) is
\begin{eqnarray}
\fl \chi_{\alpha\beta}(\bq) 
	= \chi^0_{\alpha\beta}(\bq)  
	- \sum_{i,j=1}^{19} \sum_{\ell,\ell^\prime = 0}^1 \sum_{\sigma,\sigma^\prime = \pm1} X_{\alpha}^{i\ell\sigma} (\bq) \left \{
	\left [{\bf 1}+{\tilde {\bf V}^\rho(\bq)} \tilde \chi_0(\bq)\right ]^{-1}{ \tilde {\bf V_\rho}(\bq)}
		\right \}^{i\ell\sigma,j\ell^\prime\sigma^\prime}
	X_{\beta}^{j\ell^\prime\sigma^\prime} (\bq), \nonumber \\
	\\
\fl X^{i\ell\sigma} _{\alpha}(\bq)  	
= -\frac{1}{2N} \sum_{\bk} \sum_{\theta, \theta^\prime}
\sum_{\ell_1,\ell_2} G^\sigma_{\theta^\prime \ell_1,\alpha \ell_2}(\bk) G^\sigma_{\alpha^\prime \ell_2,\theta\, \ell_1+\ell}(\bk+\bq) g^i_{\theta \theta^\prime}(\bk + \ell_1\bQ) \\
%
\fl
\tilde \chi^{i\ell\sigma,j\ell^\prime\sigma^\prime}(\bq,i\nu) = -\delta_{\sigma,\sigma^\prime}\frac{T}{2N}\sum_n \sum_{\ell_1,
  \ell_2}  \sum_{\bk}
\sum_{\mu\mu^\prime \nu \nu^\prime} g^i_{\mu^\prime\mu}(\bk +
\ell_1\bQ) G^\sigma_{\mu^\prime \ell_1 + \ell, \nu
  \ell_2+\ell^\prime}(\bk+\bq,i\omega_n +i\nu) \nonumber \\  \times G^\sigma_{\nu^\prime
  \ell_2, \mu \ell_1}(\bk,i\omega_n) g^j_{\nu\nu\prime}(\bk + \ell_2\bQ)
\end{eqnarray}
where $\omega_n = (2n+1)T\pi$ are Matsubara frequencies and $[\ldots]^{-1}$ denotes a
matrix inverse, and the $\bk$-sums are over the first AFBZ, which contains $N$ $\bk$-points.
As pointed out in Ref.~\cite{Thomson:2014}, in addition to the charge response at $\bq$ there is a spin response at $\bq+\bQ$; this additional term vanishes when spins are summed over.

\subsection{Origin of the $\bq=0$ instability in the Emery model}
A comparison between the ALJP and Emery models is made in
Fig.~\ref{fig:expandedFS}.   In both models, the Cu spectral weight is large and  uniformly distributed along the Fermi surface.  The O$p_x$ spectral weight is comparatively weak, but because the charge instability involves primarily oxygen atoms, the details of the O$p_x$ spectral weight distribution are important. 

Notably, the O$p_x$ spectral weight is highly anisotropic in the Emery model and more isotropic in the ALJP model.  (The O$p_y$ spectral function $A_{p_y}(\bk,\varepsilon_F)$ is obtained by rotating
$A_{p_x}(\bk,\varepsilon_F)$ by $90^\circ$.)
As a consequence, the matrix element of the bare susceptibility
\begin{equation}
\chi^0_{xy}(\bq= 0) \sim \sum_\bk A_{p_x}(\bk,\varepsilon_F) A_{p_y}(\bk,\varepsilon_F)
\end{equation}
is strongly reduced in the Emery model (the superscript $0$ indicates the susceptibility in
the noninteracting limit).  As we show below, this matrix element tends to stabilize the system against
nematic order.

We focus on the nonmagnetic case where approximate analytic expressions
are easily obtained.  Within a simplified random phase approximation in which
all interactions except $V_{pp}$ are ignored, we have at $\bq={\bf 0}$
\begin{equation}
\chi^\mathrm{RPA}_{3\times 3} = \left \{ 1+ \chi^0_\mathrm{3\times 3}
\left [ \begin{array}{ccc}
0 & 0 & 0 \\
0 & 0 & 8V_{pp} \\
0 &8 V_{pp} & 0 \end{array} \right ] \right \}^{-1} \chi^0_\mathrm{3\times 3},
\end{equation}
 which has a diverging eigenvalue when
\begin{equation}
 1+8V_{pp}\left[ \chi^0_{xy}-\sqrt{\chi^0_{xx}\chi^0_{yy}} 
\right] =0.
\end{equation}
(The factor of 8 arises because of a sum over spin and over the four neighboring oxygen sites
for each O$p$ orbital.)
From this equation, it is clear that $\chi^0_{xx}$ and $\chi^0_{yy}$ drive the nematic transition
while $\chi^0_{xy}$ opposes it.
Thus, it appears that the strong anisotropy of oxygen spectral weight in the Emery model is
the principal difference between the Emery and ALJP models which makes the former 
unstable to a $\bq=0$ nematic instability.

\bibliographystyle{iopart-num} 
\bibliography{RPA}

\end{document}